\documentclass{llncs}

\usepackage{amssymb, amsmath}
\usepackage{array}
\usepackage{pgfplots}
\usepackage{tikz}
\usetikzlibrary{positioning,shapes,decorations.markings,arrows,backgrounds,patterns}
\usepackage{marginnote}
\usepackage{algorithm}
\usepackage{algpseudocode}
\usepackage{ifthen}
\usepackage{todonotes}
\usepackage[hidelinks]{hyperref}
\usepackage{graphicx}
\usepackage{subcaption}
\usepackage{wrapfig}

\iffalse % color
  \graphicspath{{img-color/}}
\else % grayscale
  \selectcolormodel{gray}
  \graphicspath{{img-gray/}}
\fi

\title{Incremental Verification of Parametric and Reconfigurable Markov Chains}
\author{Paul Gainer \and Ernst Moritz Hahn \and Sven Schewe}
\institute{University of Liverpool, UK\\
\email{\{P.Gainer,E.M.Hahn,Sven.Schewe\}@liverpool.ac.uk}
}

\renewcommand{\pmatrix}{\mathbf{P}}
\newcommand{\states}{\mathcal{S}}
\newcommand{\state}{s}
\newcommand{\estate}{\state_e}
\newcommand{\init}{s_0}
\newcommand{\vars}{V}

\newcommand{\ordering}{\prec}
\newcommand{\pmc}{\mathcal{D}}
\newcommand{\pre}{\mathsf{pre}}
\newcommand{\post}{\mathsf{post}}

\newcommand{\vol}{\mathsf{Vol}}
\newcommand{\nat}{\mathbb{N}}
\newcommand{\reals}{\mathbb{R}}

\newcommand{\poly}{g}
\newcommand{\recchar}{R}
\newcommand{\recon}{\mathcal{D}^{R}}
\newcommand{\rstates}{\states^\recchar}

\newcommand{\rpmatrix}{\pmatrix^\recchar}
\newcommand{\cons}{\mathsf{Con}}
\newcommand{\recs}{\mathsf{Rec}}
\newcommand{\ints}{\mathsf{Int}}
\newcommand{\target}{\state_t}

\newcommand{\rationalfunctions}{\mathcal{F}_\vars}
\newcommand{\map}[1]{m_{#1}}
\newcommand{\domain}{\mathsf{Dom}}
\newcommand{\elim}{\mathsf{Eliminate}}
\newcommand{\infected}{\mathsf{Infected}}
\newcommand{\stateelim}{\mathsf{StateElimination}}
\newcommand{\elimset}{\mathsf{Elim}}
\newcommand{\neigh}{\mathsf{Neigh}}
\newcommand{\reach}[1]{\mathsf{reach}_{#1}}
\newcommand{\graph}{\mathcal{G}}
\newcommand{\pmrm}{\mathcal{R}}
\newcommand{\zeroconf}{Z}
\newcommand{\zinit}{\mathrm{i}}
\newcommand{\zok}{\mathrm{ok}}
\newcommand{\zerr}{\mathrm{err}}
\newcommand{\rvariable}{X}
\newcommand{\rfunc}{r}
\newcommand{\expect}{E}
\newcommand{\paths}{\mathsf{Paths}}
\newcommand{\pmeasure}{\mathsf{Pr}}
\renewcommand{\path}{\omega}

\begin{document}

\maketitle

\begin{abstract}
The analysis of parametrised systems is a growing field in verification, but the analysis of parametrised probabilistic systems is still in its infancy.
This is partly because it is much harder:
while there are beautiful cut-off results for non-stochastic systems that allow to focus only on small instances, there is little hope that such approaches extend to the quantitative analysis of probabilistic systems, as the probabilities depend on the size of a system.
The unicorn would be an automatic transformation of a parametrised system into a formula, which allows to plot, say, the likelihood to reach a goal or the expected costs to do so, against the parameters of a system.
While such analysis exists for narrow classes of systems, such as waiting queues, we aim both lower---stepwise exploring the parameter space---and higher---considering general systems.

The novelty is to heavily exploit the similarity between instances of parametrised systems.
When the parameter grows, the system for the smaller parameter is, broadly speaking, present in the larger system.
We use this observation to guide the elegant state-elimination method for parametric Markov chains in such a way, that the model transformations will start with those parts of the system that are stable under increasing the parameter.
We argue that this can lead to a very cheap iterative way to analyse parametric systems, show how this approach extends to reconfigurable systems, and demonstrate on two benchmarks that this approach scales.
\end{abstract}

\section{Introduction}
\label{sec:introduction}

Probabilistic systems are everywhere, and their analysis can be quite challenging.
Challenges, however, come in many flavours.
They range from theoretical questions, such as decidability and complexity, through algorithms design
and tool development, to the application of parametric systems.
This paper is motivated by the latter, but melds the different flavours together. 

We take our motivation from the first author's work on biologically inspired
synchronisation protocols~\cite{gainer2017investigating,gainer2017power}.
This application leaning work faced the problem of exploring a parameter space for
a family of network coordination protocols, where interacting nodes achieve consensus on
their local clocks by imitating the behaviour of fireflies~\cite{mirollo1990synchronization}.
Global clock synchronisation
emerges from local interactions between the nodes, whose behaviour is that of
coupled limit-cycle oscillators. The method used was the same that we have seen applied
by several practitioners from engineering and biology:
adjust the parameters, produce a model, and use a tool like
ePMC/IscasMC~\cite{hahn2014iscas}, PRISM \cite{kwiatkowska2011prism}, or
Storm~\cite{dehnert2017storm} to analyse it.

In the case of the synchronisation protocols, the parameters investigated
were typical of those considered when evaluating the emergence of synchronisation
in a network of connected nodes: the number of nodes forming the network, the
granularity of the model (discrete length of an oscillation cycle), the strength
of coupling between the oscillators, the likelihood of interactions between
nodes being inhibited by some external factor, for instance message loss in
a communication medium, and the length of the refractory period, an
initial period in the oscillation cycle of a node where interactions with other
nodes are ignored.

The reason to explore the parameter space can be manifold.
Depending on the application, one might simply want to obtain a feeling of how the parameters impact on the behaviour.
Another motivation is to see how the model behaves, compare it with observations, and adjust it when it fails to comply.
Regardless of the reason to adjust the parameter, the changes often lead to very similar models.

Now, if we want to analyse hundreds of similar models, then it becomes paramount to re-use as much of the analysis as possible.
With this in mind, we have selected model checking techniques for safety and reachability properties of Markov chains that build on repeated state elimination \cite{hahn2011probabilistic} as the backbone of our verification technique.
State elimination is a technique that successively changes the model by removing states.
It works like the transformation from finite automata to regular expressions:
a state is removed, and the new structure has all successors of this state as (potentially new) successors of the predecessors of this state, with the respectively adjusted probabilities
(and, if applicable, expected rewards).

If these models are changed in shape and size when playing with the parameters, then these changes tend to be smooth:
small changes of the parameters lead to small changes of the model.
Moreover, the areas that change---and, consequentially, the areas that do \emph{not} change---are usually quite easy to predict,
be it by an automated analysis of the model or by the knowledge of the expert playing with her model, 
who would know full well which parts do or do not change when increasing a parameter.
These insights inform the order in which the states are eliminated.

\begin{figure}[tb]
	\begin{center}
		\begin{subfigure}[b]{1\textwidth}\scalebox{0.8}{
			\begin{tikzpicture}
				\tikzstyle{nodestyle} = [draw, shape = circle,
						inner sep = 0pt, minimum size = 0.75cm];
				\tikzstyle{arrow} = [-stealth, thick];
				\tikzstyle{abovelabel} = [pos = 0.5, above];
				\tikzstyle{belowlabel} = [pos = 0.35, below];
				\def \spacing {1.3cm}
				\def \bendangle {30}
				\def \labelshift {(0.6, 0.1)}
				\node (init) [nodestyle] {$\zinit$};
				\node (foo) [above = 0.3cm of init] {};
				\node (ok) [nodestyle, left = \spacing of init] {$\zok$};
				\node (gap) [shape = circle, inner sep = 0pt, minimum size = 0.75cm,
						right = \spacing of init] {};
				\node (k) [nodestyle, right = \spacing of gap] {$k$};
				\node (km1) [nodestyle, right = \spacing of k] {$k{-}1$};
				\node (dots) [right = \spacing of km1] {$\ldots$};
				\node (1) [nodestyle, right = \spacing of dots] {$1$};
				\node (err) [nodestyle, right = \spacing of 1] {$\zerr$};
				\draw [arrow] (foo)--(init);
				\draw [arrow] (init)--(ok) node [abovelabel] {$1-q$};
				\draw [arrow] (init)--(k) node [abovelabel] {$q$};
				\draw [arrow] (k)--(km1) node [abovelabel] {$p$};
				\draw [arrow] (km1)--(dots) node [abovelabel] {$p$};
				\draw [arrow] (dots)--(1) node [abovelabel] {$p$};
				\draw [arrow] (1)--(err) node [abovelabel] {$p$};
				\draw [arrow] (k) to [bend left = \bendangle] node
						[belowlabel, shift={\labelshift}] {$1-p$} (init);
				\draw [arrow] (km1) to [bend left = \bendangle] node
						[belowlabel, shift={\labelshift}] {$1-p$} (init);
				\draw [arrow] (1) to [bend left = \bendangle] node
						[belowlabel, shift={\labelshift}] {$1-p$} (init);
			\end{tikzpicture}}
			\caption{}
			\label{fig:zeroconfk}
		\end{subfigure}
		\begin{subfigure}[b]{1\textwidth}\scalebox{0.8}{
			\begin{tikzpicture}
				\tikzstyle{nodestyle} = [draw, shape = circle,
						inner sep = 0pt, minimum size = 0.75cm];
				\tikzstyle{arrow} = [-stealth, thick];
				\tikzstyle{abovelabel} = [pos = 0.5, above];
				\tikzstyle{belowlabel} = [pos = 0.35, below];
				\def \spacing {1.3cm}
				\def \bendangle {30}
				\def \labelshift {(0.6, 0.1)}
				\node (init) [nodestyle] {$\zinit$};
				\node (foo) [above = 0.3cm of init] {};				
				\node (ok) [nodestyle, left = \spacing of init] {$\zok$};
				\node (kp1) [nodestyle, dashed, right = \spacing of init] {$k{+}1$};
				\node (k) [nodestyle, right = \spacing of kp1] {$k$};
				\node (km1) [nodestyle, right = \spacing of k] {$k{-}1$};
				\node (dots) [right = \spacing of km1] {$\ldots$};
				\node (1) [nodestyle, right = \spacing of dots] {$1$};
				\node (err) [nodestyle, right = \spacing of 1] {$\zerr$};
				\draw [arrow] (foo)--(init);				
				\draw [arrow] (init)--(ok) node [abovelabel] {$1-q$};
				\draw [arrow, dashed] (init)--(kp1) node [abovelabel] {$q$};
				\draw [arrow, dashed] (kp1)--(k) node [abovelabel] {$p$};				
				\draw [arrow] (k)--(km1) node [abovelabel] {$p$};
				\draw [arrow] (km1)--(dots) node [abovelabel] {$p$};
				\draw [arrow] (dots)--(1) node [abovelabel] {$p$};
				\draw [arrow] (1)--(err) node [abovelabel] {$p$};
				\draw [arrow, dashed] (kp1) to [bend left = \bendangle] node
						[belowlabel, shift={\labelshift}] {$1-p$} (init);
				\draw [arrow] (k) to [bend left = \bendangle] node
						[belowlabel, shift={\labelshift}] {$1-p$} (init);
				\draw [arrow] (km1) to [bend left = \bendangle] node
						[belowlabel, shift={\labelshift}] {$1-p$} (init);
				\draw [arrow] (1) to [bend left = \bendangle] node
						[belowlabel, shift={\labelshift}] {$1-p$} (init);
			\end{tikzpicture}}
			\caption{}
			\label{fig:zeroconfkp1}
		\end{subfigure}
	\end{center}
	\caption{The Zeroconf Protocol for $n = k$ (a) and $n = k+1$ (b).}
	\label{fig:zeroconf}
\end{figure}
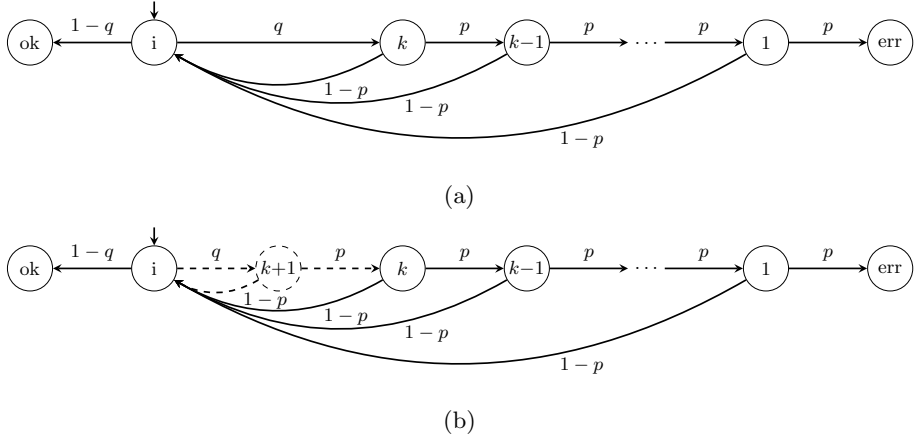

When, say, the increase of a parameter allows for re-using all elimination steps but the last two or three, then repeating the analysis is quite cheap.
Luckily, this is typically the case in structured models, e.g.\ those who take a chain-, ring-, or tree-like shape.
As a running example of a structured model we consider the Zeroconf protocol~\cite{bohnenkamp2002cost}
for the autonomous configuration of multiple hosts in a network with
unique IP addresses (Figure \ref{fig:zeroconf}).
A host that joins the network selects an address uniformly at random from
$a$ available addresses. If the network consists of $h$ hosts, then probability that
the selected address is already in use is $q = \frac{h}{a}$.

The protocol then checks $n$ times if the selected address is already in use by sending a request to the network.
If the address is fresh (which happens with a probability of $1-q$), none of these tests will fail, and the address will be classed as new.
If the address is already in use (which happens with a probability of $q$), then each test is faulty: collisions go undetected with some likelihood $p$ due to message loss and time-outs.
When a collision is detected (which happens with a likelihood of $1-p$ in each attempt, provided that a collision exists), then
the host restarts the process.
If a collision has gone undected after $n$ attempts, then the host will
incorrectly assume that its address is unique.

While the family of Zeroconf protocols is also parameterised in the transition probabilities,
we are mostly interested in their parametrisation in the
structure of the model. Figures~\ref{fig:zeroconfk} and~\ref{fig:zeroconfkp1}
show the models for $n=k$ and $n=k+1$, respectively, successive checks after each selection of an IP.
These two models are quite similar: they structurally differ only in the introduction of a single
state, and the transitions that come with this additional state.
If we are interested in
calculating the function that represents the probability of reaching the
state $\mathsf{err}$ in both models, where this function is given in terms
of individual rational functions that label the transitions,
then the structural similarities allow us to re-use the intermediate
terms obtained from the evaluation for $n=k$ when evaluating for $n=k+1$.

The structure of the paper is as follows. We begin by comparing our work to similar approaches
in Section~\ref{sec:relatedwork}.
In Section~\ref{sec:algorithms}, we introduce the novel algorithms for the
analysis of reconfigured models.
We then evaluate our approach on
two different types of parametric models which are discussed in
Section~\ref{sec:casestudies}. Section~\ref{sec:conclusion} concludes the
paper and outlines future work.

\section{Related Work}
\label{sec:relatedwork}
Our work builds on previous results in the area of parametric Markov model checking and incremental runtime verification of stochastic systems.

Daws~\cite{daws2004symbolic} considered Markov chains, which are parametric in the transition probabilities, but not in their graph structure.
He introduced an algorithm to calculate the function that represents the probability of reaching a set of target states for all well-defined evaluations for a parametric Markov chain.
For this, he interprets the Markov chain under consideration as a finite automaton, in which transitions are labelled with symbols that correspond to rational numbers or variables.
He then uses state elimination~\cite{hopcroft2008introduction} to obtain a regular expression for the language of the automaton.
Evaluating these regular expressions into rational functions yields the probability of reaching the target states.
One limiting factor of this approach is that the complete regular expression has to be stored in memory.

Hahn et al. introduced~\cite{hahn2011probabilistic} and implemented \cite{hahn2010param} a simplification and refinement of Daws' algorithm.
Instead of using regular expressions, they store rational functions directly.
This has the advantage that possible simplifications of these functions, such as cancellation of common factors, can be applied on the fly.
This allows memory to be saved. It also provides a more concise representation of of the values computed to the user.
They have also extended the scope of the approach from reachability, to additionally handle accumulated reward properties.
Several works from RWTH Aachen have followed up on solution methods for parametric Markov chains~\cite{DehnertJJCVBKA15,JansenCVWAKB14,QuatmannD0JK16}.
This type of parametric model checking has been used in~\cite{BartocciGKRS11} to build a model-repair framework for stochastic systems and in~\cite{JohnsonK11,JohnsonK12,Johnson15} to reason about the robustness of robot controllers against sensor errors.

Our paper borrows some ideas from the work of Kwiatkowska et al.~\cite{kwiatkowska2011incremental}. Their work considers MDPs that are labelled with parametric transition probabilities.
The authors do not aim to compute a closed-form function that represents properties of a model, but rather at accelerating the computation of results for individual instantiations of parameter values.
Rather than state elimination, they use value iteration and other solution methods to evaluate the model for certain parameter values.
In doing so, they can for instance, re-use computations for different instantiations of parameters that only depend on the graph structure of the model that remains unchanged for different instantiations.

We also take inspiration from Forejt et al.~\cite{forejt2012incremental}, where
the role of parameters is different.
Forejt et al.\ describe a policy iteration-based technique to evaluate parametric MDPs.
While they also considered parameters in~\cite{forejt2012incremental} that can influence the model structure, they would exploit similarities to inform the search for the policy when moving from one parameter value to the next.
The repeatedly called model checking of Markov chains, on the other hand, is not parametric.
Our approach is therefore completely orthogonal, as we focus on the analysis of Markov chains.
In more detail, Forejt et al.~\cite{forejt2012incremental} would use an incremental approach to find a good starting point for a policy iteration approach for MDPs.
The intuition there is that an optimal policy is likely to be good---if not optimal---on the shared part of an MDP that grows with a parameter.
This approach has the potential to find the policy in less steps, as less noise disturbs the search in smaller MDPs, but its main promise is to provide an excellent oracle for a starting policy.
Moreover, in the lucky instances where the policy is stable, it can also happen that there is a part of the Markov chain, obtained by using a policy that builds on a smaller parameter, that is outside of the cone of influence of the changes to the model.
In this case, not only the policy, but also its evaluation is stable under the parameter change.

%Delete: The assumption of \cite{forejt2012incremental} is that parameters only influence the validity of \emph{guards}.
%This way, from one parameter instantiation to the next, one probability distribution to be nondeterministically chosen in a given state can appear or disappear, but probability distributions do not change.
%In our approach parameters do influence probability distributions - due to lack of nondeterminism, there is always exactly one possible probability distribution per state.

\section{Algorithms}
\label{sec:algorithms}

We first describe the state elimination method of
Hahn~\cite{hahn2011probabilistic} for parametric
Markov Chains (PMCs), and then
introduce an algorithm that substantially reduces the
cost of recomputation of the parametric
reachability probability for a reconfigured PMC. First we
give some general definitions.
Given a function $f$ we denote the domain of $f$ by
$\domain(f)$. We use the notation
$f \oplus f^\prime = f \restriction_{\domain(f)
		\setminus \domain(f^\prime)} \cup f^\prime$
to denote the overriding union of $f$ and $f^\prime$.
Let $\vars = \{v_1, \ldots, v_n\}$ denote a set of variables
over~$\reals$.
%An evaluation for $\vars$ is a partial
%function $\evaluation : \vars \to \reals$.
A \emph{polynomial} $\poly$ over $\vars$ is a sum of monomials
\begin{align*}
	\poly(v_1, \ldots, v_n) = \sum_{i_1, \ldots, i_n}
			a_{i_1},\ldots ,_{i_n} v_1^{i_1} \ldots v_n^{i_n},
\end{align*}
where each $i_j \in \nat$ and each $a_{i_1},\ldots,_{i_n} \in \reals$.
A \emph{rational function} $f$ over a set of variables $\vars$
is a fraction
$f(v_1,\ldots,v_n) = \frac{f_1(v_1,\ldots,v_n)}{f_2(v_1,\ldots,v_n)}$
of two polynomials $f_1, f_2$ over $\vars$. We denote the set
of rational functions from $\vars$ to $\reals$ by $\rationalfunctions$.

\begin{definition}
A \emph{parametric Markov chain} (PMC) is a tuple
$\pmc = (\states, \init, \pmatrix, \vars)$, where $\states$
is a finite set of states, $\init$ is the initial state,
$\vars = \{v_1, \ldots, v_n\}$ is a finite set of parameters,
and $\pmatrix$ is the probability matrix
$\pmatrix : \states \times \states \to \rationalfunctions$.
\end{definition}

A \emph{path} $\path$ of a PMC $\pmc = (\states, \init, \pmatrix, \vars)$
is a non-empty finite, or infinite, sequence $\state_0, \state_1, \state_2, \ldots$
where $\state_i \in \states$ and $\pmatrix(\state_i, \state_{i+1}) > 0$ for $i \geqslant 0$.
We denote the $i^{th}$ state of $\path$ by $\path[i]$, 
the set of all paths starting in state $\state$ by $\paths(\state)$, and the
set of all finite paths starting in $\state$ by $\paths_f(\state)$.
For a finite
path $\path_f \in \paths_f(\state)$ the \emph{cylinder set} of $\path_f$
is the set of all infinite paths in $\paths(\state)$ that share the
prefix $\path_f$.
The probability of taking a finite
path $\state_0, \state_1, \ldots, \state_n \in \paths_f(\state_0)$
is given by $\prod_{i=1}^n \pmatrix(\state_{i-1}, \state_i)$.
This measure over finite paths can be extended to a probability measure
$\pmeasure_\state$ over the set of infinite paths $\paths(\state)$, where
the smallest $\sigma$-algebra over $\paths(\state)$ is the smallest
set containing all cylinder sets for paths in $\paths_f(\state)$. For a detailed
description of the construction of the probability measure we refer the
reader to~\cite{kemeny2012denumerable}.

\begin{definition}
Given a PMC $\pmc = (\states, \init, \pmatrix, \vars)$,
the \emph{underlying graph} of $\pmc$
is given by $\graph_\pmc = (\states, E)$ where 
$E = \{(\state, \state^\prime) \mid
		\pmatrix(\state, \state^\prime) > 0 \}$.
\end{definition}
Given a state $\state$, we denote the set
of all immediate predecessors and successors of $\state$
in the underlying graph of $\pmc$ by $\pre_\pmc(\state)$ and
$\post_\pmc(\state)$, respectively, and we define the
\emph{neighbourhood} of $\state$ as
$\neigh(\state) = \state \cup \pre_\pmc(\state) \cup \post_\pmc(\state)$.
We write $\reach{D}(\state, \state^\prime)$ if
$\state^\prime$ is reachable from $\state$ in the underlying graph
of $D$. 

\subsection{State Elimination}
\label{subsec:stateelim}
The algorithm of Hahn~\cite{hahn2011probabilistic} proceeds as follows,
where the input is a PMC $\pmc = (\states, \init, \pmatrix, \vars)$ and a
set of target states $T \subset \states$. Initially, preprocessing is applied
and without loss of generality all outgoing transitions from states in
$T$ are removed and a new state $\target$ is introduced such that
$\pmatrix(t, \state_t) = 1$ for all $t \in T$. All states $s$,
where $s$ is unreachable from the initial state or $T$ is unreachable
from $s$, are then removed along with all incident transitions. A
state $\estate$ in $\states \setminus \{\init, \target\}$ is then chosen for
elimination and Algorithm~\ref{alg:stateelim} is applied. Firstly, for every
pair $(\state_1, \state_2) \in \pre_\pmc(\estate) \times \post_\pmc(\estate)$,
the existing probability $\pmatrix(\state_1, \state_2)$ is incremented
by the probability of reaching $\state_2$ from $\state_1$ via $\estate$.
The state and any incident transitions are then eliminated from $D$.
This  procedure is repeated until only $\init$ and $\target$ remain,
and the probability of reaching $T$ from $\init$ is then given by
$\pmatrix(\init, \target)$.

\begin{algorithm}[tb]
	\caption{State Elimination}
	\label{alg:stateelim}
	\begin{algorithmic}[1]
		\Procedure{\textsc{StateElimination}}{$\pmc, \estate$}
		\State{\textbf{requires:}}
		A PMC $\pmc$ and $\estate$, a state to eliminate in $\pmc$.
		\ForAll{$(\state_1, \state_2) \in \pre_\pmc(\estate)
				\times \post_\pmc(\estate)$}
			\State $\pmatrix(\state_1, \state_2) \gets
					\pmatrix(\state_1, \state_2) + \pmatrix(\state_1, \estate)
					\frac{1}{1 - \pmatrix(\estate, \estate)}
					\pmatrix(\estate, \state_2)$		
		\EndFor
		\State $\elim(\pmc, \estate)$
			\textit{// remove $\estate$ and incident transitions from $\pmc$}
		\State \textbf{return} $\pmc$
		\EndProcedure
	\end{algorithmic}
\end{algorithm}

\subsection{Reconfiguration}

Recall that we are interested in the re-use of information
when recalculating reachability for a reconfigured PMC. We can do this by
choosing the order in which we eliminate states in the original
PMC. The general idea is that, if the set of states where structural changes
might occur is known a priori, then we can apply state elimination to all other
states first. We say that states where
structural changes might occur are \emph{volatile} states. 
\begin{definition}
A \emph{volatile} parametric Markov chain (VPMC) is a tuple
$\pmc = (\states, \init, \pmatrix, \vars, \vol)$ where 
$(\states, \init, \pmatrix, \vars)$ is a PMC and 
$\vol \subseteq \states$ is a set of volatile states
for $\pmc$.
\end{definition}

Given a VPMC $\pmc = (\states, \init, \pmatrix, \vars, \vol)$,
we can define an \emph{elimination ordering} for $\pmc$ as a
bijection $\ordering_\pmc: \states \to \{1, \ldots, \left|\states\right|\}$
that defines an ordering for the elimination of states in $S$, such that
${\ordering_\pmc}(s) < {\ordering_\pmc}(s^\prime)$ holds for all
$s \in \states \setminus \vol, s^\prime \in \vol$, where
${\ordering_\pmc}(s) < {\ordering_\pmc}(s^\prime)$ indicates that $\state$
is eliminated before $\state^\prime$. Observe that a volatile state
in $\pmc$ is only eliminated after all non-volatile states.

\begin{definition}
A \emph{reconfiguration} for a VPMC
$\pmc = (\states, \init, \pmatrix, \vars, \vol)$ is a PMC
$\recon = (\rstates, \init, \rpmatrix, \vars)$, where $\rstates$ is a set of
states with $\rstates \cap \states \ne \emptyset$, $\init$ and $\vars$
are the initial state and the finite set of parameters for $\pmc$.
The reconfigured probability matrix $\rpmatrix$ is a total function
$\rpmatrix : \rstates \times \rstates \to \rationalfunctions$
such that, for all $\state, \state^\prime \in \rstates$ where
$\pmatrix(\state, \state^\prime)$ is defined,
$\pmatrix(\state, \state^\prime) \ne \rpmatrix(\state, \state^\prime)$
implies $\state, \state^\prime \in \vol$.
\end{definition}

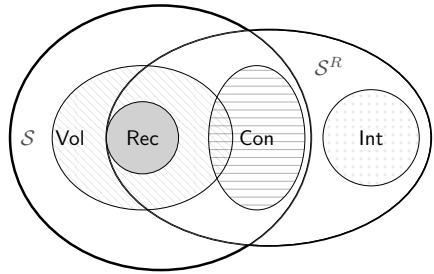
\begin{wrapfigure}{r}{0.5\textwidth}
	\vspace{-15pt}
	\begin{center}
		\def\statescircle{(0,0) ellipse (2.5 and 2.2)}
		\def\rstatescircle{(1.8,0) ellipse (2.7 and 1.8)}
		\def\volatilecircle{(-0.3, 0) ellipse (1.5 and 1.2)}
		\def\reconfcircle{(-0.3, 0) ellipse (0.6 and 0.6)}
		\def\constcircle{(1.6,0) ellipse (0.8 and 1.2)}
		\def\intcircle{(3.5,0) ellipse (0.8 and 0.8)}
		\scalebox{0.8}{
		  \begin{tikzpicture}[legend/.style={draw, circle, minimum size = 0.4cm}]
			\node[] at (-2.2, 0) {$\states$};
			\node[] at (2.8, 1.2) {$\rstates$};
			\fill[white, opacity=0.3] \statescircle;
			\draw[very thick] \statescircle;
			\fill[white, opacity=0.3] \rstatescircle;
			\draw[thick] \rstatescircle;
			\begin{scope}
				\clip \rstatescircle;
				\fill[white] \statescircle;
				%\fill[pattern=dots, opacity=0.3] \statescircle;
				\draw[thick] \rstatescircle;
			\end{scope}
			\draw \statescircle;
			%\fill[white] \volatilecircle;
			\fill[pattern=north west lines, opacity=0.3] \volatilecircle;
			\draw \volatilecircle;
			\fill[white] \reconfcircle;			
			\fill[color={rgb:red,0.1;green,0.1;blue,0.1}, opacity=0.27] \reconfcircle;
			\draw \reconfcircle;
			\fill[pattern=horizontal lines, opacity=0.3] \constcircle;
			\draw \constcircle;
			\fill[pattern=dots, opacity=0.3] \intcircle;			
			\draw \intcircle;
			
			\node[] at (-1.5, 0) {$\vol$};
%			\node[] at (-0.45, 0) {$\vol$};
			\node[] at (1.6, 0) {$\cons$};
			\node[] at (-0.3, 0) {$\recs$};
			\node[] at (3.5, 0) {$\ints$};

%			\matrix[] at (3.8, -1.8) {
%				\node[legend, very thick,
%					label=right:{$\states$}]  {}; \\
%				\node[legend, dashed,
%					label=right:{$\rstates$}]  {}; \\
%			};
%			\matrix[draw] at (2.5, 2.5) {
%				\node[legend, pattern=north west lines,
%					fill opacity=0.3,
%					label=right:{$\scriptstyle\vol$}]  {}; \\
%				\node[legend, fill={rgb:red,0.1;green,0.1;blue,0.1},
%					fill opacity=0.27,
%					label=right:{$\scriptstyle\recs$}]  {}; \\
%				\node[legend, label=right:{$\scriptstyle\cons$}]  {}; \\
%				\node[legend, pattern=dots,
%					fill opacity=0.3,
%					label=right:{$\scriptstyle\ints$}]  {}; \\				
%			};
\iffalse
			            \draw[rounded corners=10] (3,1.5) ellipse (3 and 1.5); % set S
                        \node at (0.3,1.5) {$S$};
                        \draw[rounded corners=10] (6,1.5) ellipse (3 and 1.3); % set S^R
                        \node at (7.8,1.5) {$S^R$};
                        \draw[rounded corners=10,dashed] (2.5,1.5) ellipse (1.8 and 1.2); % set Vol
                        \node at (1.1,1.5) {$\mathit{Vol}$};
                        \draw[rounded corners=10,dotted] (2.7,1.5) ellipse (3.8 and 2.6); % set Rec
                        \node at (2.5,1.5) {$\mathit{Rec}$};
                        
                        \draw[rounded corners=10,dotted] (4,0.4) rectangle (5.7,2.6);
                        \node at (5.15,1.5) {$\mathit{Con}$};
                        \draw[rounded corners=10,dashed] (6.2,0.4) rectangle (7.3,2.2);
                        \node at (6.7,1.5) {$\mathit{Int}$};
\fi
		\end{tikzpicture}}
		\caption{Venn diagram showing the consistent, reconfigured, and
		introduced states for a VPMC $\pmc$ and reconfiguration $\recon$.}
		\label{fig:venndiagram}
	\end{center}
	\vspace{-20pt}
\end{wrapfigure}
Given a VPMC $\pmc$ and a reconfiguration $\recon$ for $\pmc$
we say that a state $\state$ in $\states$ is \emph{consistent}
in $\recon$ if $\state$ is also in $\rstates$, 
the set of all predecessors and
successors of $\state$ remains unchanged in $\recon$,
(that is $\pre_\pmc(\state) = \pre_{\recon}(\state)$,
$\post_\pmc(\state) = \post_{\recon}(\state)$,
$\pmatrix(\state_1, \state) = \rpmatrix(\state_1, \state)$
for every $\state_1 \in \pre_\pmc(\state)$), and
$\pmatrix(\state, \state_2) = \rpmatrix(\state, \state_2)$
for every $\state_2 \in \post_\pmc(\state)$. We say that a state $\state$ in
$\states$ is \emph{reconfigured} in $\recon$ if $\state$ is also in
$\states^\prime$ and $\state$ is not consistent. Finally, we say
that a state $\state$ in $\states^\prime$ is \emph{introduced} in $\recon$ if
$\state$ is neither consistent nor reconfigured. We 
denote the sets of all consistent, reconfigured, and introduced states
by $\cons(\pmc, \recon), \recs(\pmc, \recon)$, and $\ints(\pmc, \recon)$,
respectively. Figure~\ref{fig:venndiagram} shows the consistent, reconfigured, and
introduced states for $\pmc$ and $\recon$.
%, where $\cons(\pmc, \recon) \cap \recs(\pmc, \recon){=}\emptyset$, and $\recs(\pmc, \recon) \subseteq \vol$.

Algorithm~\ref{alg:buildmap} computes the %function representing the
parametric reachability probability of some target state in
a VPMC $\pmc = (\states, \init, \pmatrix, \vars, \vol)$
for a given elimination ordering for $\pmc$.
Observe that we compute
the reachability probability with respect to a single target state.
The reachability of a set of target states can be computed
by first removing all outgoing transitions from
existing target states, and then introducing a new target state to
which a transition is taken from any existing target state with
probability $1$. We introduce a new initial
state to the model, from which a transition is taken to the original
initial state with probability $1$.
The algorithm computes a % function mapping
partial probability matrix $\pmatrix^\prime$, initialised as a
zero matrix, that stores the probability of
reaching $\state_2$ from $\state_1$ via any eliminated
non-volatile state, where $\state_1, \state_2$ are either volatile
states, the initial state, or the target state.
It also computes an \emph{elimination map}, a function mapping 
tuples of the form $(\estate, \state_1, \state_2)$, where $\estate$ is
an eliminated volatile state and $\state_1, \state_2$ are either volatile
states, the initial state, or the target state, to the value computed
during state elimination for the probability of reaching $\state_2$
from $\state_1$ via $\estate$. We are only interested in transitions
between volatile states, the initial state, or the target state,
since all non-volatile states in any reconfiguration of $\pmc$ will
be eliminated first. Computed values for transitions
to or from these states therefore serve no purpose once they have been eliminated.

\begin{algorithm}[tb]
  \caption{Parametric Reachability Probability for VMPCs}
	\label{alg:buildmap}
	\begin{algorithmic}[1]
		\Procedure{\textsc{ParametricReachability}}{$\pmc$, $\ordering_\pmc$,
		$\target$}
		\State{\textbf{requires:}}
		a target state $\target \in \states$,
		and for all $\state \in \states$ it holds
		$\reach{\pmc}(\init, \state)$ and
		$\reach{\pmc}(\state, \target)$.
		\State $E \gets \states \setminus \{\init, \target\}$
			\qquad \textit{// states to be eliminated from} $\pmc$
		%\State $\map{\pmc} \gets \emptyset$ \mhcommentalg{same here for $\map{\pmc}$}
		\State $\pmatrix^\prime \gets 0_{\lvert \states \rvert, \lvert \states \rvert}$
			\hspace{1.05cm} \textit{// partial probability matrix}
		\State $\map{\pmc}^\vol \gets \emptyset$
			\hspace{0.8cm} \qquad \textit{// elimination map }
		\While{$E \ne \emptyset$}
			\State $\estate \gets \arg \min \ordering_\pmc \restriction_E$
			\ForAll{$(\state_1, \state_2) \in \pre_\pmc(\estate)
					\times \post_\pmc(\estate)$}
				\State $p = \pmatrix(\state_1, \estate)
						\frac{1}{1 - \pmatrix(\estate, \estate)}
						\pmatrix(\estate, \state_2)$
				\If{$\state_1 \in \vol \cup \{\init, \target\} \text{ and }
						\state_2 \in \vol \cup \{\init, \target\}$}
					\If{$\estate \not \in \vol$}
						\State $\pmatrix^\prime(\state_1, \state_2) \gets
								\pmatrix^\prime(\state_1, \state_2) + p$
					\Else					
						\State $\map{\pmc}^\vol \gets \map{\pmc}^\vol
								\oplus \{(\estate, \state_1, \state_2) \mapsto p\}$
					\EndIf
				\EndIf					
				\State $\pmatrix(\state_1, \state_2) \gets
						\pmatrix(\state_1, \state_2) + p$	
			\EndFor
			\State $\elim(\pmc, \estate)$ 
				\textit{// remove $\estate$ and incident transitions from $\pmc$}
			\State $E \gets E \setminus \{\estate\}$
		\EndWhile
		\State \textbf{return} $(\pmatrix(\init, \target), \pmatrix^\prime, \map{\pmc}^\vol)$
		\EndProcedure
	\end{algorithmic}
\end{algorithm}

Given a reconfiguration $\recon = (\rstates, \init, \rpmatrix, \vars)$
for $\pmc$, an elimination ordering for $\pmc$, and the partial
probability matrix and mapping computed
using Algorithm~\ref{alg:buildmap},
Algorithm~\ref{alg:reconfigure} computes the parametric reachability
probability for $\recon$ as follows. Firstly the set of all non-volatile
states of $\pmc$ are eliminated in $\recon$, though state
elimination itself does not occur. A set of \emph{infected} states
is then initialised to be the set of all states that are reconfigured in
$\recon$. Then, for every other remaining state that is not introduced
in $\recon$, if that state or its neighbours are not infected we treat this
state as a non-volatile state. That is, we update $\pmatrix^\prime$ with
the corresponding values in $\map{\pmc}^\vol$ and eliminate the state without
performing state elimination. If the state, or one of its neighbours,
is infected then the probability matrix is updated such that all
transitions to and from that state are augmented with the
corresponding values in $\pmatrix^\prime$. These entries are then removed from
the mapping. Subsequently, state elimination (Algorithm~\ref{alg:stateelim})
is  applied, and the infected area is expanded to include the immediate
neighbourhood of the eliminated state. Finally, state elimination is
applied to the set of all remaining introduced states in $\recon$.

%\pgcomment{Need to update the map and partial probability matrix here
%and return them too.}{}
\begin{algorithm}[tb]
	\caption{Parametric Reachability Probability for reconfigured VMPC}
	\label{alg:reconfigure}
	\begin{algorithmic}[1]
		\Procedure{\textsc{ReconfiguredParametricReachability}}
			{$\pmc$, $\recon$, $\ordering_\pmc$, $\pmatrix^\prime$,
				$\map{\pmc}^\vol$, $\target$}
		\State{\textbf{requires:}}
%		VPMC $\pmc = (\states, \init, \pmatrix, \vars, \vol)$,
%		a reconfiguration $\recon = (\rstates, \init, \rpmatrix, \vars)$
%		for $\pmc$, an elimination
%		ordering $\ordering_\pmc$, partial probability matrix $\pmatrix^\prime$
%		and elimination map $\map{\pmc}^\vol$ for $\pmc$,
		absorbing target state $\target$ such that $\target \in \states$ and
		$\target \in \rstates$, for all 
		$\state \in \states$ it holds	$\reach{\recon}(\init, \state)$ and
		$\reach{\recon}(\state, \target)$, and for all 
		$\state^\prime \in \rstates$ it holds $\reach{\recon}(\init, \state^\prime)$ and
		$\reach{\recon}(\state^\prime, \target)$.
%		\State{\textbf{Require:}}
%		VPMC $\pmc = (\states, \init, \pmatrix, \vars, \vol)$,
%		a reconfiguration $\recon = (\rstates, \init, \rpmatrix, \vars)$
%		for $\pmc$, an elimination
%		ordering $\ordering_\pmc$, partial probability matrix $\pmatrix^\prime$
%		and elimination map $\map{\pmc}^\vol$ for $\pmc$,
%		an absorbing target state $\target$ such that $\target \in \states$ and
%		$\target \in \rstates$, for all 
%		$\state \in \states$ it holds	$\reach{\recon}(\init, \state)$ and
%		$\reach{\recon}(\state, \target)$, and for all 
%		$\state^\prime \in \rstates$ it holds $\reach{\recon}(\init, \state^\prime)$ and
%		$\reach{\recon}(\state^\prime, \target)$.
		\State $M \gets (\vol \cap \rstates) \cup \{\init, \target\}$
		\State $\elimset \gets \cons(\pmc, \recon) \setminus M$
		\State $\elim(\recon, \elimset)$ 
				\textit{// remove all $\estate \in \elimset$ and incident transitions from $\pmc$}		
		%\mhcommentalg{what does this call do again exactly, does it just remove states and their transitions?.}
		\State $\elimset \gets \vol \cap \rstates$
		\State $\infected \gets \recs(\pmc, \recon)$
		\State $\rpmatrix(\init, \target) = \pmatrix^\prime(\init, \target)$ 
		\While{$\elimset \ne \emptyset$}
			\State $\estate \gets \arg \min \ordering_\pmc
				\restriction_{\elimset}$
			\If{$\infected \cap \neigh(\estate) = \emptyset$}
				\ForAll{$(\estate^\prime, \state_1, \state_2) \in
						\domain(\map{\pmc}^\vol \restriction_{\{\estate\} \times M^2})$}
					\State $\pmatrix^\prime(\state_1, \state_2)
						\gets \pmatrix^\prime(\state_1, \state_2) +
						\map{\pmc}^\vol(\estate^\prime, \state_1, \state_2)$
				\EndFor
				\State $\elim(\recon, \estate)$
					\textit{// remove $\estate$ and incident transitions from $\recon$}				
			\Else
				\ForAll{$\{(\state_1, \state_2) \in \rstates \times \rstates \mid \state_1 = \estate \text{ or } \state_2 = \estate\}$}
					\State $\rpmatrix(\state_1, \state_2) \gets
								\rpmatrix(\state_1, \state_2) +
								\pmatrix^\prime(\state_1, \state_2)$
					\State $\pmatrix^\prime(\state_1, \state_2) \gets 0$
%					\State $\map{\pmc}(\state_1, \state_2) \gets \map{\pmc}(\state_1, \state_2)
%							\restriction_{\domain(\map{\pmc}) \setminus
%								\{(\state_1, \state_2)\}}$  \mhcommentalg{what does this line do?}
				\EndFor
				\State $\recon \gets \stateelim (\recon, \estate)$
				\State $\infected \gets \infected \cup \neigh(\estate)$
			\EndIf
			\State $\elimset \gets \elimset \setminus \{\estate\}$
		\EndWhile
7		\ForAll{$\estate \in \ints(\pmc, \recon)$}
			\State $\recon \gets \stateelim (\recon, \estate)$
		\EndFor
		\State \textbf{return} $\rpmatrix(\init, \target)$
		\EndProcedure
	\end{algorithmic}
\end{algorithm}

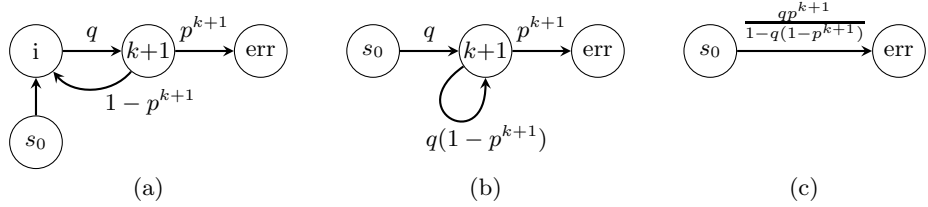
\begin{figure}[tb]
	\begin{center}
			\begin{tikzpicture}
				\tikzstyle{nodestyle} = [draw, shape = circle,
						inner sep = 0pt, minimum size = 0.7cm];
				\tikzstyle{arrow} = [-stealth, thick];
				\tikzstyle{abovelabel} = [pos = 0.5, above];
				\tikzstyle{belowlabel} = [pos = 0.35, below];
				\def \spacing {0.78cm}
				\def \bendangle {50}
				\def \labelshift {(0.6, 0.1)}
				\node (init) [nodestyle] {$\zinit$};
				\node (s0) [nodestyle, below = 0.5cm of init] {$\init$};
				\node (kp1) [nodestyle, right = \spacing of init] {$k{+}1$};
				\node (err) [nodestyle, right = \spacing of kp1] {$\zerr$};
				\draw [arrow] (s0)--(init);				
				\draw [arrow] (init)--(kp1) node [abovelabel] {$q$};
				\draw [arrow] (kp1)--(err) node [abovelabel] {$p^{k+1}$};
				\draw [arrow] (kp1) to [bend left = \bendangle] node
						[belowlabel, shift={\labelshift}] {$1-p^{k+1}$} (init);
				\node (s02) [nodestyle, right = \spacing of err] {$\init$};
				\node (kp12) [nodestyle, right = \spacing of s02] {$k{+}1$};
				\node (err2) [nodestyle, right = \spacing of kp12] {$\zerr$};
				\draw [arrow] (s02)--(kp12) node [abovelabel] {$q$};
				\draw [arrow] (kp12)--(err2) node [abovelabel] {$p^{k+1}$};
				\draw [arrow] (kp12)  to[in = 270,out = 215, loop] node
						[belowlabel, shift={(0.6, -0.1)}] {$q (1-p^{k+1})$} (kp12);
				\node (s03) [nodestyle, right = \spacing of err2] {$\init$};
				\node (err3) [nodestyle, right = 2.3*\spacing of s03] {$\zerr$};
				\draw [arrow] (s03)--(err3) node [pos = 0.5, above] {$\frac{q p^{k+1}}{1-q (1-p^{k+1})}$};
				\node (label1) [below = 1.2cm of kp1] {(a)};
				\node (label2) [below = 1.2cm of kp12] {(b)};
				\node (label3) [right = 4.6 * \spacing of label2] {(c)};
			\end{tikzpicture}
	\end{center}
	\caption{$\zeroconf_{k+1}$ after the elimination of states $k$ (a), $i$ (b), 
	and $k+1$ (c).}
	\label{fig:zeroconfelim}
\end{figure}

\begin{example}
\label{ex:zeroconf}
Consider again the Zeroconf models from Figures~\ref{fig:zeroconfk}
and~\ref{fig:zeroconfkp1}. Let
$\zeroconf_k = (\states, \init, \pmatrix, \vars, \vol)$ be a VPMC for
$n = k$, such that $\states = \{1, \ldots, k\} \cup \{\init, \zinit, \zerr\}$,
$\vars = \{p, q\}$, and $\vol = \{\zinit, k\}$. We are
interested in the parametric reachability probability of the state
$\zerr$. Note that preprocessing removes the state $\zok$ from $\zeroconf_k$
since $\reach{\zeroconf_k}(\zok, \zerr)$ does not hold. Now define
$\ordering_{\zeroconf_k} = \{1 \mapsto 1, 2 \mapsto 2, \ldots,
		k \mapsto k, \zinit \mapsto k+1\}$ to be an elimination
ordering for $\zeroconf_k$. State elimination then proceeds according
to $\ordering_{\zeroconf_k}$, and after the first $k-1$ states
have been eliminated we have
\begin{align*}
	\pmatrix^\prime(k, \zerr) = p^k, \quad
	\pmatrix^\prime(k, \zinit) = \sum_{i=1}^{k-1} (p^j - p^{j+1}). \\
%	\map{\zeroconf_k} = \{
%		(k, \zerr) \mapsto p^k, 
%		(k, \zinit) \mapsto \sum_{i=1}^{k-1} (p^j - p^{j+1})
%	\}.
\end{align*}
%\pgcomment{We can probably omit this part since $\vol$ and
%$\recs(\zeroconf_k, \zeroconf_{k+1})$ coincide, and hence
%$\map{\zeroconf_k}^\vol$ is never used.}{}
Eliminating the remaining volatile states $k$ and $\zinit$ then
yields
\begin{align*}
	\map{\zeroconf_k}^\vol = \{
		(k, \zinit, \zerr) \mapsto q p^k,
		(k, \zinit, \zinit) \mapsto q (1-p^k),
		(\zinit, \init, \zerr) \mapsto \frac{q p^k}{1-q (1-p^k)}
	\}.
\end{align*}

Now let $\zeroconf_{k+1} = (\rstates, \init, \rpmatrix, \vars)$ be a
reconfiguration for $\zeroconf_k$ such that $\rstates = \states \cup \{k+1\}$.
We then have
$\cons(\zeroconf_k, \zeroconf_{k+1}) = \{1 \ldots k-1\} \cup \{\init, \zerr\}$,
$\recs(\zeroconf_k, \zeroconf_{k+1}) = \{k, i\}$,
and $\ints(\zeroconf_k, \zeroconf_{k+1}) = \{k+1\}$.
First, all states $1, \ldots, k-1$ and their incident transitions are
simply eliminated from $\zeroconf_{k+1}$, and the infected set is
initialised to be $\{k, i\}$. Since $k$ is already infected
we update the probability matrix as follows, 
\begin{align*}
	\rpmatrix(k, \zerr) & \gets \rpmatrix(k, \zerr) + \pmatrix^\prime(k, \zerr) \\
	& \gets 0 + p^k = p^k,  \\
	\rpmatrix(k, \zinit) & \gets \rpmatrix(k, \zinit){+}\pmatrix^\prime(k, \zinit) \\
	& \gets (1{-}p){+} \sum_{i=1}^{k{-}1} (p^j{-}p^{j{+}1}){=}\sum_{i=0}^{k-1} (p^j{-}p^{j+1}) = 1{-}p^k.
\end{align*}
State elimination is then applied to state $k$ and the corresponding entries in
$\pmatrix^\prime$ are set to zero. The state of the model after this step is shown
in Figure~\ref{fig:zeroconfelim}a. State $\zinit$ is also infected, but this
time there are no corresponding non-zero values in $\pmatrix^\prime$. State elimination
is then applied to state $\zinit$ resulting in the model shown in
Figure~\ref{fig:zeroconfelim}b. Finally, state elimination is applied to the
single introduced state $k+1$, resulting in the model shown in
Figure~\ref{fig:zeroconfelim}c, and the algorithm terminates.
\end{example}

\subsection{Correctness}
The correctness of the approach follows as an easy corollary from the correctness of Hahn's general state elimination approach~\cite{hahn2011probabilistic}.
We outline the simple inductive argument, starting with the first parameter under consideration---which serves as the induction basis---and then look at incrementing the parameter value---which serves as the induction step.

For the induction basis, the first parameter considered, there is really nothing to show:
we would merely choose a particular order in which states are eliminated, and the correctness of Hahn's state-elimination approach does not depend on the order in which states are eliminated.

For the induction step, consider that we have an order for one parameter value, and that we have an execution of the state elimination along this given order $<_o$.
Our approach then builds a new order for the next parameter value.
The new order $<_n$ is quite closely linked to the old order $<_o$, but for correctness, a very weak property suffices.

To prepare our argument, let us consider a set $E$ of states with the following properties:
the neighbourhood of $E$ is the same in the Markov chains for the old and new parameter;
the restriction of $<_o$ and $<_n$ to $E$ define the same order; and
$E$ is the set of smallest states w.r.t.\ $<_o$ and $<_n$ ($s \in E$ and $(s' <_o s \vee s' <_n s)$ implies $s' \in E$).
In this case, the initial sequence of the first $|E|$ reductions for the new Markov chain (along $<_n$) are the same as the first $|E|$ state eliminations along the old Markov chain (along $<_o$).
Consequently, these elimination steps can be re-used, rather than re-done.

In Algorithm~\ref{alg:reconfigure} we require less: we still require that the neighbourhood of $E$ is the same in the Markov chains for the old and new parameter and the restriction of $<_o$ and $<_n$ to $E$ define the same order, but relax the third requirement to
$s \in E \mbox{ and } (s' <_o s \vee s' <_n s) \mbox{ implies that } s' \in E \mbox { or } s' \mbox{ is no neighbour of } s.$
The result is the same:
for the states in $E$, the $|E|$ state eliminations for the new Markov chain (along $<_n$) are the same as $|E|$ state eliminations along the old Markov chain (along $<_o$).
Consequently, these elimination steps can be re-used.

\subsection{Extension to Parametric Markov Reward Models}

We now describe how we can extend the algorithms to PMCs annotated with rewards.

\begin{definition}
A Parametric Markov Reward Model (PMRM) is a tuple
$\pmrm = (\pmc, \rfunc)$ where $\pmc = (\states, \init, \pmatrix, \vars)$
is a PMC and $\rfunc : \states \to \rationalfunctions$ is the reward function.
\end{definition}
The reward function labels states in $\pmrm$ with a rational function
over $\vars$ that corresponds to the reward that is gained if that state
is visited.
Given a PMRM $\pmrm = (\pmc, \rfunc)$ with
$\pmc = (\states, \init, \pmatrix, \vars)$, we are interested in the
\emph{parametric expected accumulated reward}~\cite{kwiatkowska2007stochastic}
until some target state $\target \in \states$ is reached. This is defined
as the expectation of the random variable
$\rvariable^\pmrm : \paths(\state_0) \to \reals \cup \{\infty\}$
over the infinite paths of $\pmrm$. Given the set
$\path_{\target} = \{i \mid w[i] = \target\}$ we define
\begin{align*}
\rvariable^\pmrm(\path) =
\begin{cases}
	\begin{array}{ll}
	\infty	&
		\text{   if } \path_{\target} = \emptyset \\
	\sum_{i=0}^{k-1} r(\path[i]) &
		\text{   otherwise, where } k = \min \path_{\target},
	\end{array}	
\end{cases}
\end{align*}
and define the expectation of $\rvariable^\pmrm$ with respect to
$\pmeasure_{\init}$ as
\begin{align*}
\expect[\rvariable^\pmrm] = 
	\sum_{\path \in \paths(\init)} \rvariable^\pmrm(\path) \pmeasure_{\init}(\path).
\end{align*}

We extend our notion of volatility to PMRMs as follows. We say that a state
is volatile if structural changes might occur in that state \emph{or} if the reward
labelling that state might change. Because of space limitations we omit the
full definitions for volatile PMRMs, but the constructions are straightforward.
Algorithms~\ref{alg:stateelim} to~\ref{alg:reconfigure} are extended to
incorporate rewards. For Algorithm~\ref{alg:stateelim}, in
addition to updating the probability matrix for the elimination of some state
$\estate$, we also update the reward function as follows,
\begin{align*}
\rfunc(\state_1) \gets \rfunc(\state_1) + \pmatrix(\state_1, \estate) 
	\frac{\pmatrix(\estate, \estate)}{1 - \pmatrix(\estate, \estate)}
	\rfunc(\estate).
%\rfunc(\state_2) \gets \rfunc(\state_2) + 
%	\frac{\pmatrix(\state_1, \estate) \pmatrix(\estate, \state_2)}
%				{1 - \pmatrix(\estate, \estate)}
%	\rfunc(\estate).
\end{align*}
The updated value for $\rfunc(\state_1)$ reflects the reward that would be
accumulated if a transition would be taken from $\state_1$ to $\estate$, where
the expected number of self transitions would be
$\frac{\pmatrix(\estate, \estate)}{1 - \pmatrix(\estate, \estate)}$.
%Note that this value is independent of the successors of se. 
Algorithm~\ref{alg:buildmap} then constructs additional mappings to
record these computed expected reward values, which are then used for
reconfiguration in Algorithm~\ref{alg:reconfigure}.

\section{Case Studies}
\label{sec:casestudies}

We provide a prototypical implementation\footnote{\url{https://github.com/PaulGainer/PMC}} of the
technique and
define the metric that we will use for the evaluation of
different models to be the total number of arithmetic
operations performed for the elimination of all states in a model.
Our implementation serves only to illustrate the potential of the
method, and we will integrate the technique into the
probabilistic model checker ePMC~\cite{hahn2014iscas}. 

%\pgcomment{If we don't get to 16 pages, remove this}{}
Due to space limitations we restrict our analysis to two classes
of models. Firstly we consider the family of Zeroconf protocols
described in Section~\ref{sec:introduction}, and secondly we
consider a family of models used for the analysis of biologically
inspired firefly synchronisation protocols---the class of
protocols that inspired this work. 

\subsection{Zeroconf}

We are interested in the reachabiliy of the error state for
the family of Zeroconf models, parameterised in the number $n$ of attempts,
after which the protocol will (potentially incorrectly) assume that it has selected a unique address.
The initial model for $n=1$ is defined, 
its volatile region is determined as in Example~\ref{ex:zeroconf},
and Algorithm~\ref{alg:buildmap} is applied. In each incremental
step we increment $n$ and apply Algorithm~\ref{alg:reconfigure} to the model.
Volatile states can be identified in each step.

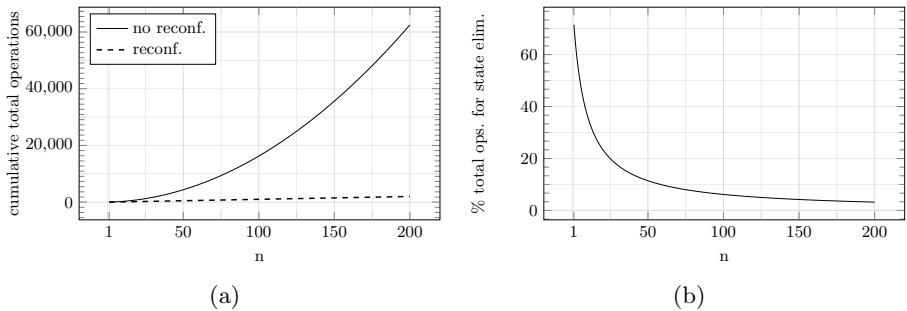
\begin{figure}[tb]
	\begin{subfigure}[t]{0.5\textwidth}
		\renewcommand{\axisdefaultheight}{160pt}
		\scalebox{0.7}{
		\begin{tikzpicture}
			\begin{axis}[scaled ticks = false, minor y tick num=10,
					legend style={at={(0.03, 0.85)},anchor=west},
					legend cell align={left}, xlabel={n},
					ylabel={cumulative total operations}, xtick={1,50,100,150,200},
					extra x ticks={0,25,50,75,100,125,150,175,200},
					extra y ticks={0,10000,30000,50000,60000},
					extra x tick labels={},
					extra y tick labels={},
					tick style={grid=major},
					grid style={opacity=0.3},
					extra tick style={grid=major,opacity=0.2}]
				\addplot[smooth, no markers] table [x = n, y = op, col sep = comma]
								 {zeroconf.csv};
				%\addplot table [x = n, y = op, col sep = comma, each nth point = 10] {zeroconf.csv};
				\addplot[smooth, no markers, dashed, thick]
						table [x = n, y = reop, col sep = comma] {zeroconf.csv};
				%\addplot table [x = n, y = reop, col sep = comma, each nth point = 10] {zeroconf.csv};
				\addlegendentry{no reconf.}
				\addlegendentry{reconf.}
			\end{axis}
		\end{tikzpicture}}
		\caption{}
		\label{fig:zeroconfcase}
	\end{subfigure}
	\begin{subfigure}[t]{0.5\textwidth}
		\renewcommand{\axisdefaultheight}{160pt}
		\scalebox{0.7}{
		\begin{tikzpicture}
			\begin{axis}[scaled ticks = false, minor y tick num=5, xlabel={n},
					ylabel={\% total ops. for state elim.}, xtick={1,50,100,150,200},
					extra x ticks={0,25,50,75,100,125,150,175,200},
					extra y ticks={0,10,30,50,70},
					extra x tick labels={},
					extra y tick labels={},
					tick style={grid=major},
					grid style={opacity=0.3},
					extra tick style={grid=major,opacity=0.2}]
				\addplot[smooth, no markers]
						table [x = n, y = perc, col sep = comma] {zeroconf.csv};
				%\addplot table [x = n, y = perc, col sep = comma, each nth point = 10] {zeroconf.csv};
			\end{axis}
		\end{tikzpicture}}
		\caption{}
		\label{fig:zeroconfcaseratio}
	\end{subfigure}
	\caption{Cumulative total of arithmetic operations performed for iterative analysis of Zeroconf
	for $n = 1\ldots 200$ (a), and the ratio of total operations for reconfiguration to
	total operations for regular state elimination, given as a percentage~(b).}
\end{figure}

Figures~\ref{fig:zeroconfcase} and~\ref{fig:zeroconfcaseratio}
show the total number of performed arithmetic operations 
accumulated during the incremental analysis of the models and the ratio
of the number of arithmetic operations performed for regular state elimination, respectively.
This ratio shows the small share of the number of iterations required when the values are calculated for a range of parameters in our approach
(repeated applications of Algorithm~\ref{alg:reconfigure}), when compared to the na\"ive approach to re-calculate all values from scratch (applying Algorithm~\ref{alg:buildmap}).

Figure~\ref{fig:zeroconfcase} shows that
the total number of operations is quadratic in the parameter when regular state elimination (applying Algorithm~\ref{alg:buildmap}) is repeatedly applied from scratch.
This is a consequence of the number of operations for \emph{each} parameter being linear in the paramter value when na\"ively applying Algorithm~\ref{alg:buildmap}.
This is in stark contrast to the number of operations needed when the parameter is stepwise incremented using Algorithm~\ref{alg:reconfigure}, stepwise capitalising on the analysis of the respective predecessor model.
Here the update cost is constant: since the extent of structural change at each
step is constant.
This leads to dramatic savings (quadratic vs.\ linear) when exploring the parameter space, as illustrated by Figure~\ref{fig:zeroconfcaseratio}.

\subsection{Oscillator Synchronisation}

We now consider the models developed
in~\cite{gainer2017investigating} and~\cite{gainer2017power} to analyse protocols
for the clock synchronisation of nodes in a network.
In these protocols, consensus
on clock values emerges from interactions between the nodes.
The underlying mathematical model is that of coupled oscillators.
This family of models is parametric in 
the number $N$ of nodes that form the network;
the granularity $T$ of the discretisation of the oscillation cycle; the length  $R$ of the refractory
period, during which nodes ignores interactions with their neighbours;
the strength $\epsilon$ of the coupling between the oscillators; and
finally the likelihood $\mu$ of any individual interaction between two nodes not occurring due to some external factor. 

Each state of the model corresponds to some global configuration for the
network---a vector encoding the size of node clusters that share the same progress
through their oscillation cycle. The target states of interest are those
in which all nodes share the same progression through their cycle and
are therefore \emph{synchronised}.

Changing the parameters $N$ and $T$ redefines the encoding of a global network state.
This results in drastic changes to the structure of the model and therefore 
makes it hard to identify volatile states.
Our prototypical
implementation only considers low-level models defined explicitly as a set of
states and a transition matrix, which trivialises the identification of volatile areas.
Future implementation into
ePMC, however, will allow volatile states to be clearly identified by analysing the guards
present in high-level model description languages~\cite{alur1999reactive}.
This works in particular for the parameters $N$ and $T$ we have studied.

Changing the parameter $\epsilon$ results in such severe changes in the
structure of the model that we do not see how the synergistic effects we
have observed can be ported to analysing its parameter space,
while changing $\mu$ does not change the structure of the
underlying graph and hence is not interesting for what we want to show. 

In this paper, we therefore focus on the incremental analysis for the parameter $R$.
We arbitrarily fix $N$ to be $5$ and $\epsilon$ to be $0.1$, and repeat the
incremental analysis for four different values for $T$. The parameter $R$ varies from $1$ to $T$ (for each of the different values of $T$ we have considered).

\begin{figure}[tb]
	\begin{subfigure}[t]{0.5\textwidth}
		\begin{tabular}{cc}
		\scalebox{0.7}		
		{
		\begin{tikzpicture}[scale=0.5,font=\Large]
			\begin{axis}[scaled ticks = false, minor y tick num=5, legend style={at={(0.95, 0.45)},anchor=east}, legend cell align={left}, xlabel={R}, ylabel={cumulative total operations}, title={T=4}, xtick distance = 1,
					tick style={grid=major},
					grid style={opacity=0.3}]
				\addplot table [x = R, y = op, col sep = comma, each nth point = 1] {synchT4.csv};
				\addplot table [x = R, y = reop, col sep = comma, each nth point = 1] {synchT4.csv};
				\addlegendentry{no reconf.}
				\addlegendentry{reconf.}
			\end{axis}
		\end{tikzpicture}} &
		\scalebox{0.7}		
		{
		\begin{tikzpicture}[scale=0.5,font=\Large]
			\begin{axis}[scaled ticks = false, minor y tick num=5, legend style={at={(0.03, 0.85)},anchor=west}, legend cell align={left}, xlabel={R}, title={T=5}, xtick distance = 1,
					tick style={grid=major},
					grid style={opacity=0.3}]
				\addplot table [x = R, y = op, col sep = comma, each nth point = 1] {synchT5.csv};
				\addplot table [x = R, y = reop, col sep = comma, each nth point = 1] {synchT5.csv};
			\end{axis}
		\end{tikzpicture}} \\	
\scalebox{0.7}		
		{
		\begin{tikzpicture}[scale=0.5,font=\Large]
			\begin{axis}[scaled ticks = true, minor y tick num=5, legend style={at={(0.95, 0.45)},anchor=east}, legend cell align={left}, xlabel={R}, ylabel={}, title={T=6}, xtick distance = 1,
					tick style={grid=major},
					grid style={opacity=0.3}]
				\addplot table [x = R, y = op, col sep = comma, each nth point = 1] {synchT6.csv};
				\addplot table [x = R, y = reop, col sep = comma, each nth point = 1] {synchT6.csv};
			\end{axis}
		\end{tikzpicture}} &
		\scalebox{0.7}		
		{
		\begin{tikzpicture}[scale=0.5,font=\Large]
			\begin{axis}[scaled ticks = true, minor y tick num=5, legend style={at={(0.03, 0.85)},anchor=west}, legend cell align={left}, xlabel={R}, title={T=7}, xtick distance = 1,
					tick style={grid=major},
					grid style={opacity=0.3}]
				\addplot table [x = R, y = op, col sep = comma, each nth point = 1] {synchT7.csv};
				\addplot table [x = R, y = reop, col sep = comma, each nth point = 1] {synchT7.csv};
			\end{axis}
		\end{tikzpicture}} \\	
		\end{tabular}
		\caption{}
		\label{fig:synchcase}
	\end{subfigure}
	\begin{subfigure}[t]{0.5\textwidth}
		\begin{tabular}{cc}
		\scalebox{0.7}		
		{
		\begin{tikzpicture}[scale=0.5,font=\Large]
			\begin{axis}[scaled ticks = false, minor y tick num=5, xlabel={R}, ylabel={\% total ops. for state elim.}, title={T=4}, xtick distance = 1,
					tick style={grid=major},
					grid style={opacity=0.3}]			
				\addplot table [x = R, y = perc, col sep = comma, each nth point = 1] {synchT4.csv};
			\end{axis}
		\end{tikzpicture}} &
		\scalebox{0.7}		
		{
		\begin{tikzpicture}[scale=0.5,font=\Large]
			\begin{axis}[scaled ticks = false, minor y tick num=5, xlabel={R}, title={T=5}, xtick distance = 1,
					tick style={grid=major},
					grid style={opacity=0.3}]			
				\addplot table [x = R, y = perc, col sep = comma, each nth point = 1] {synchT5.csv};
			\end{axis}
		\end{tikzpicture}} \\	
		\scalebox{0.7}		
		{
		\begin{tikzpicture}[scale=0.5,font=\Large]
			\begin{axis}[scaled ticks = false, minor y tick num=5, xlabel={R}, ylabel={\vphantom{a}}, title={T=6}, xtick distance = 1,
					tick style={grid=major},
					grid style={opacity=0.3}]			
				\addplot table [x = R, y = perc, col sep = comma, each nth point = 1] {synchT6.csv};
			\end{axis}
		\end{tikzpicture}} &
		\scalebox{0.7}		
		{
		\begin{tikzpicture}[scale=0.5,font=\Large]
			\begin{axis}[scaled ticks = false, minor y tick num=5, xlabel={R}, title={T=7}, xtick distance = 1,
					tick style={grid=major},
					grid style={opacity=0.3}]			
				\addplot table [x = R, y = perc, col sep = comma, each nth point = 1] {synchT7.csv};
			\end{axis}
		\end{tikzpicture}} \\	
		\end{tabular}
		\caption{}
		\label{fig:synchcaseratio}
	\end{subfigure}
	\caption{Cumulative total of arithmetic operations performed for iterative analysis of
	synchronisation models with respect to $R$ (a), and the ratio of totals
	for reconfiguration to 	totals for regular state elimination given as a percentage (b).}
\end{figure}
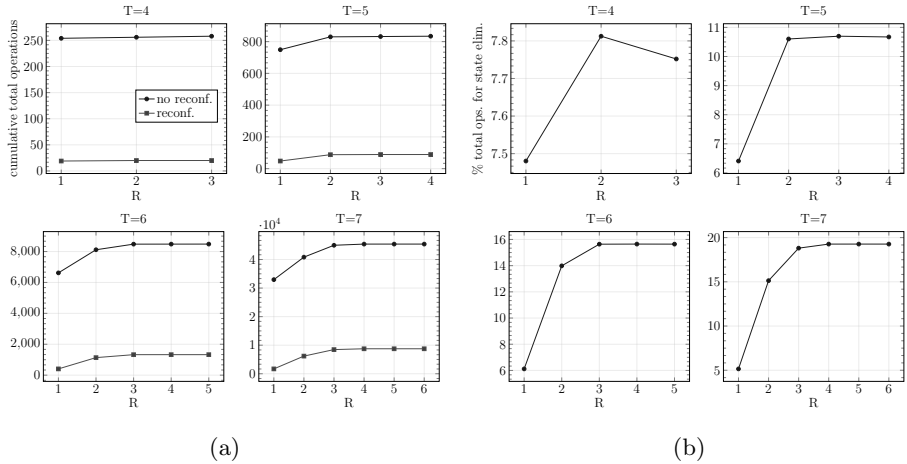

Figures~\ref{fig:synchcase} and~\ref{fig:synchcaseratio}
show the total number of performed arithmetic operations and the ratio
of the totals for regular state elimination to the totals for
reconfiguration given as a percentage, respectively.
The effectiveness of the
approach lessens as $T$ increases, a result of the rounding of
real values to discrete integer values that occurs when generating
the transitions for the initial model~\cite{gainer2017investigating}.
Higher values of $T$ result in an increase in the number of possible
successor states for global states of the network, which in turn leads to an
increase in the number of transitions in the model. Similarly, incrementing
$R$ results in reduced effectiveness as fewer interactions between
nodes are ignored, and again more transitions are introduced to the model.

Overall it is clear that, while still substantial, the gains here are not
as pronounced as those seen for the analysis of the Zeroconf protocol.
This is to be expected, since the structural changes induced by changing
the parameter $R$ are not constant for each iteration---the higher
the value of $R$ the greater the extent of the structural changes incurred.
\section{Conclusion and Future Work}
\label{sec:conclusion}

It is clear---and, in hindsight, unsurprising---that our approach works well for structured
Markov chains, such as chain-, ring-, or tree-like structures.
Our experiments have lent evidence to this by showing that that where the cost of
model-checking an individual model grows linearly with a parameter, model checking up
to a parameter becomes linear in the maximal parameter considered,
whereas the overall costs grow quadratically if all models are considered individually.
Thus, we expect significant gain wherever changes can be localised and isolated.
Moreover, we expect this to be the norm rather than the exception.
After all, chains, rings, and trees are common structures in models.

It is quite striking that \emph{very} specialised structures
have enjoyed a lot of attention, and so have \emph{absolutely} general ones.
The standard example for very specialised structures is waiting queues.
Fixed length waiting queues, for example, have closed form solutions.
Thus, when the system analyst creates a structure, which is so standard that it has
a known closed form solution \emph{and}---and this is a big `and'---realises that
this is the case and looks up the closed form solution, then this analysis is the unicorn.
However, if the structure is slightly different, if she fails to see that the problem
has a closed form solution, or if she does not want to invest the time to research the
closed form solution, then she would currently have to fall back to the na\"ive solution.
Here our technique is a nice sweet spot between these extremes: the speed is close to
evaluating closed form solutions, but applying our method does not put any burden on the system analyst
who creates the parametrised model.

The limitations of our model are that it loses much of its advantage when a change
in a parameter induces severe structural changes in the model. For the synchronisation
protocol, some parameters severely change the structure. % include the granularity of the
%discretisation of the oscillation cycle.
This is because most of the nodes are connected
by an edge, and for such dense graphs, structural changes can have a huge cone of influence.

The next step of our work will be to tap the full potential of our approach by integrating
it into the probabilistic model checker ePMC \cite{hahn2014iscas}. Here the
symbolic description of the system will expose the volatile areas
and---more importantly---the non-volatile areas that appear to be stable under
successive increments of the parameter values.
We also expect to obtain synergies by combining our method with the approach of \cite{forejt2012incremental}, extending our approach to models with non-determinism, such as interactive Markov chains and Markov decision processes.

\section{Acknowledgements}

This work was supported by the Sir Joseph Rotblat Alumni Scholarship at Liverpool,
EPSRC grants EP/M027287/1 and EP/N007565/1, and by the Marie Sk{\l}odowska Curie Fellowship \emph{Parametrised Verification and Control}.

\bibliographystyle{splncs03}
\bibliography{literature}

\begin{thebibliography}{10}
\providecommand{\url}[1]{\texttt{#1}}
\providecommand{\urlprefix}{URL }

\bibitem{alur1999reactive}
Alur, R., Henzinger, T.A.: Reactive modules. Formal methods in system design
  15(1),  7--48 (1999)

\bibitem{BartocciGKRS11}
Bartocci, E., Grosu, R., Katsaros, P., Ramakrishnan, C.R., Smolka, S.A.: Model
  repair for probabilistic systems. In: TACAS. pp. 326--340 (2011)

\bibitem{bohnenkamp2002cost}
Bohnenkamp, H., {van der Stok}, P., Hermanns, H., Vaandrager, F.:
  Cost-optimization of the IPv4 zeroconf protocol, pp. 531--540. IEEE Computer
  Society Press (2003)

\bibitem{daws2004symbolic}
Daws, C.: Symbolic and parametric model checking of discrete-time markov
  chains. In: International Colloquium on Theoretical Aspects of Computing. pp.
  280--294. Springer (2004)

\bibitem{DehnertJJCVBKA15}
Dehnert, C., Junges, S., Jansen, N., Corzilius, F., Volk, M., Bruintjes, H.,
  Katoen, J., {\'{A}}brah{\'{a}}m, E.: {PROPhESY}: {A} probabilistic parameter
  synthesis tool. In: CAV. pp. 214--231 (2015)

\bibitem{dehnert2017storm}
Dehnert, C., Junges, S., Katoen, J.P., Volk, M.: A storm is coming: A modern
  probabilistic model checker. In: CAV. pp. 592--600. Springer (2017)

\bibitem{forejt2012incremental}
Forejt, V., Kwiatkowska, M., Parker, D., Qu, H., Ujma, M.: Incremental runtime
  verification of probabilistic systems. In: International Conference on
  Runtime Verification. pp. 314--319. Springer (2012)

\bibitem{gainer2017investigating}
Gainer, P., Linker, S., Dixon, C., Hustadt, U., Fisher, M.: Investigating
  parametric influence on discrete synchronisation protocols using quantitative
  model checking. In: QEST. pp. 224--239. Springer (2017)

\bibitem{gainer2017power}
Gainer, P., Linker, S., Dixon, C., Hustadt, U., Fisher, M.: The power of
  synchronisation: Formal analysis of power consumption in networks of
  pulse-coupled oscillators. arXiv preprint arXiv:1709.04385  (2017)

\bibitem{hahn2010param}
Hahn, E.M., Hermanns, H., Wachter, B., Zhang, L.: Param: A model checker for
  parametric markov models. In: CAV. pp. 660--664. Springer (2010)

\bibitem{hahn2011probabilistic}
Hahn, E.M., Hermanns, H., Zhang, L.: Probabilistic reachability for parametric
  markov models. STTT  13(1),  3--19 (2011)

\bibitem{hahn2014iscas}
Hahn, E.M., Li, Y., Schewe, S., Turrini, A., Zhang, L.: iscas m c: a web-based
  probabilistic model checker. In: FM. pp. 312--317. Springer (2014)

\bibitem{hopcroft2008introduction}
Hopcroft, J.E.: Introduction to automata theory, languages, and computation.
  Pearson Education India (2008)

\bibitem{JansenCVWAKB14}
Jansen, N., Corzilius, F., Volk, M., Wimmer, R., {\'{A}}brah{\'{a}}m, E.,
  Katoen, J., Becker, B.: Accelerating parametric probabilistic verification.
  In: QEST. pp. 404--420 (2014)

\bibitem{JohnsonK11}
Johnson, B., Kress{-}Gazit, H.: Probabilistic analysis of correctness of
  high-level robot behavior with sensor error. In: Robotics: Science and
  Systems (2011)

\bibitem{JohnsonK12}
Johnson, B., Kress{-}Gazit, H.: Probabilistic guarantees for high-level robot
  behavior in the presence of sensor error. Autonomous Robots  33(3),  309--321
  (2012)

\bibitem{Johnson15}
Johnson, B.L.: Synthesis, analysis, and revision of correct-by-construction
  controllers for robots with sensing and actuation errors. Ph.D. thesis,
  Cornell University (2015)

\bibitem{kemeny2012denumerable}
Kemeny, J.G., Snell, J.L., Knapp, A.W.: Denumerable Markov chains: with a
  chapter of Markov random fields by David Griffeath, vol.~40. Springer Science
  \& Business Media (2012)

\bibitem{kwiatkowska2007stochastic}
Kwiatkowska, M., Norman, G., Parker, D.: Stochastic model checking. In:
  International School on Formal Methods for the Design of Computer,
  Communication and Software Systems. pp. 220--270. Springer (2007)

\bibitem{kwiatkowska2011prism}
Kwiatkowska, M., Norman, G., Parker, D.: Prism 4.0: Verification of
  probabilistic real-time systems. In: CAV. pp. 585--591. Springer (2011)

\bibitem{kwiatkowska2011incremental}
Kwiatkowska, M., Parker, D., Qu, H.: Incremental quantitative verification for
  markov decision processes. In: International Conference on Dependable Systems
  \& Networks. pp. 359--370. IEEE (2011)

\bibitem{mirollo1990synchronization}
Mirollo, R.E., Strogatz, S.H.: Synchronization of pulse-coupled biological
  oscillators. SIAM Journal on Applied Mathematics  50(6),  1645--1662 (1990)

\bibitem{QuatmannD0JK16}
Quatmann, T., Dehnert, C., Jansen, N., Junges, S., Katoen, J.: Parameter
  synthesis for markov models: Faster than ever. In: ATVA. pp. 50--67 (2016)

\end{thebibliography}

\end{document}